\documentclass{article}
\usepackage{amsmath}
\usepackage{amsfonts}
\usepackage{graphicx}
\usepackage{bm}
\usepackage{amsmath} 
\usepackage{array} 
\usepackage{graphicx} 
\usepackage{booktabs} 
\usepackage{lmodern} 
\usepackage{placeins}
\usepackage{caption}
\usepackage{authblk}
\usepackage{natbib}
\setcitestyle{authoryear,open={(},close={)}}

\title{Tracking the Hidden Forces Behind Laos’ 2022 Exchange Rate Crisis and Balance of Payments Instability}

\author[1]{Mariza Cooray}
\author[2]{Rolando Gonzales Martinez}
\affil[2]{University of Groningen. Contact: r.m.gonzales.martinez@rug.nl}

\date{}

\begin{document}

\maketitle

\begin{abstract}
\noindent This working paper uses a Dynamic Factor Model ('the model') to identify underlying factors contributing to the debt-induced economic crisis in the People's Democratic Republic of Laos ('Laos'). The analysis aims to use the latent macroeconomic insights to propose ways forward for forecasting. We focus on Laos's historic structural weaknesses to identify when a balance of payments crisis with either a persistent current account imbalance or rapid capital outflows would occur. By extracting latent economic factors from macroeconomic indicators, the model provides a starting point for analyzing the structural vulnerabilities leading to the value of the kip in USD terms dropping and contributing to inflation in the country. This findings of this working paper contribute to the broader literature on exchange rate instability and external sector vulnerabilities in emerging economies, offering insights on what constitutes as 'signals' as opposed to plain 'noise' from a macroeconomic forecasting standpoint.
\end{abstract}

\begin{center}
\textbf{Executive Summary}    
\end{center}

We estimated the macroeconomic factors underlying exchange rate and balance of payments (BoP) dynamics in Laos. These risk factors drive the economy, but they are not directly observable. Instead, we approximate them using observable economic indicators. Because these factors evolve over time, tracking their changes helps mitigate future crises more effectively than monitoring every individual economic variable. By summarizing the overall economic condition, these risk factors provide valuable insights for economic stability.

Our modeling results (Tables \ref{tab:XR} and \ref{tab:XRQ}) show that several negative factors, building up since 2005, contributed to the exchange rate crisis of 2022 (Figure \ref{fig:XRL}). The most significant drivers included the decline in foreign direct investment beginning in 2005 (Figure \ref{fig:FDI}), the deterioration of the current account balance, and the reduction in total reserves (as a percentage of debt) since 2010 (Figures \ref{fig:CAB} and \ref{fig:RPD}). Additionally, the increasing debt service burden, evident since 2015 (Figure \ref{fig:DS}), further exacerbated the crisis.

In the short term, the worsening of the overall BoP and the surge in the emission of narrow money (M1) since 2020 intensified inflationary pressures, creating uncertainty that further depreciated the value of the kip and accelerated the exchange rate decline potentially due to "hedging" as mentioned by the World Bank in their routine Lao Economic Monitors as a result of the declining value of the Kip.

Laos' external sector accounts began showing signs of deterioration in 2016 (Figure \ref{fig:BoPL}), primarily due to the worsening of the current account’s primary income (Figure \ref{fig:CAPI_BoP}) and the capital account (Figure \ref{fig:KA_BoP}).

\section{Introduction}

Laos PDR is currently experiencing a severe economic crisis, driven by both external and internal factors, including rising debt levels, inflation, currency depreciation, and supply chain disruptions. The sharp depreciation of the Lao kip has exacerbated inflationary pressures by increasing the cost of imports, especially fuel and essential goods. As a result, the cost of living has risen, fueling social unrest. 

Inflation has also surged due to rising global food and fuel prices, compounded by Laos’ heavy reliance on imports. These challenges have made it difficult to maintain basic living standards, affecting key sectors such as healthcare, education, and employment. The COVID-19 pandemic and geopolitical tensions related to the war in Ukraine have further disrupted global supply chains, causing shortages and price hikes for essential goods. However, because the data sample concludes in 2022—the year the war began—it limits the statistical assessment of the war’s impact on Laos.

We used dynamic factor models (DFMs) to calculate underlying factors that reflect the economic conditions affecting the exchange rate and the balance of payments (BoP) in Laos. Since these underlying factors are not directly observable, we estimated them using available macroeconomic indicators at both annual and quarterly levels. DFMs applied to annual data capture medium-to-long-term trends, while those estimated with monthly or quarterly data reflect short-term shocks.

Our DFM results show that exchange rate depreciation in Laos stemmed from high external debt servicing, depleting reserves, current account imbalances, declining foreign direct investment, and Laos’ reliance on foreign currencies. The DFM applied to BoP components highlights that BoP vulnerabilities primarily resulted from a deteriorating capital account and declining primary income. However, improvements in the financial account and portfolio investment provided some stability.

\section{Methodology}

Dynamic Factor Models (DFMs) are a tool to analyze large datasets by summarizing information through a few unobserved common factors. \citet{forni2000generalized}  introduced the Generalized Dynamic Factor Model (GDFM) as a framework for identifying and estimating common factors in extensive datasets. Their methodology addressed challenges in high-dimensional settings, providing consistent estimations of the number of factors and their loadings. Later, \cite{stock2002macroeconomic} proposed the use of diffusion indexes, constructed from dynamic factor models, to forecast macroeconomic variables. They showed that these indexes effectively summarize information from large datasets, enhancing forecast accuracy. \citet{bai2008forecasting} provided a comprehensive analysis of factor models in economic forecasting, including methods for determining the number of factors, estimation techniques, and practical implementations of these models in forecasting economic time series. 

\citet{stock2011dynamic} reviewed the development and applications of dynamic factor models, its theoretical results, estimation methods, and various applications, highlighting the versatility and effectiveness of DFMs in econometric analysis. More recently, \citet{barhoumi2013dynamic} offered a comprehensive review of the literature on dynamic factor models, discussing their theoretical foundations, estimation techniques, and applications in macroeconomic forecasting. They also addressed challenges and recent advancements in the field. These studies collectively emphasize the significance of dynamic factor models for analyzing large-scale data and improving forecasting accuracy. 

A dynamic-factor model takes on the form:

\[
\mathbf{y}_t = \mathbf{P} \mathbf{f}_t + \mathbf{Q} \mathbf{x}_t + \mathbf{u}_t
\]
\[
\mathbf{f}_t = \mathbf{R} \mathbf{w}_t + \sum_{i=1}^{p} \mathbf{A}_i \mathbf{f}_{t-i} + \mathbf{\nu}_t
\]
\[
\mathbf{u}_t = \sum_{i=1}^{q} \mathbf{C}_i \mathbf{u}_{t-i} + \mathbf{\epsilon}_t
\]

where the definition of each term in the equations is given in the  table below:

\begin{table}[h!]
\centering
\begin{tabular}{ccl}
\hline
\textbf{Item} & \textbf{Dimension} & \textbf{Definition} \\ \hline
$\mathbf{y}_t$ & $k \times 1$ & vector of dependent variables \\ 
$\mathbf{P}$ & $k \times n_f$ & matrix of parameters \\ 
$\mathbf{f}_t$ & $n_f \times 1$ & vector of unobservable factors \\ 
$\mathbf{Q}$ & $k \times n_x$ & matrix of parameters \\ 
$\mathbf{x}_t$ & $n_x \times 1$ & vector of exogenous variables \\ 
$\mathbf{u}_t$ & $k \times 1$ & vector of disturbances \\ 
$\mathbf{R}$ & $n_f \times n_w$ & matrix of parameters \\ 
$\mathbf{w}_t$ & $n_w \times 1$ & vector of exogenous variables \\ 
$\mathbf{A}_i$ & $n_f \times n_f$ & matrix of autocorrelation parameters for $i = 1, \dots, p$ \\ 
$\mathbf{\nu}_t$ & $n_f \times 1$ & vector of disturbances \\ 
$\mathbf{C}_i$ & $k \times k$ & matrix of autocorrelation parameters for $i = 1, \dots, q$ \\ 
$\mathbf{\epsilon}_t$ & $k \times 1$ & vector of disturbances \\ \hline
\end{tabular}
\end{table}
\FloatBarrier
By selecting different numbers of factors and lags, the dynamic-factor model encompasses the six models in the table below:

\begin{table}[h!]
\centering
\begin{tabular}{lccc}
\hline
\textbf{Model} & $n_f$ & $p$ & $q$ \\ \hline
Dynamic factors with vector autoregressive errors & $n_f > 0$ & $p > 0$ & $q > 0$ \\ 
Dynamic factors & $n_f > 0$ & $p > 0$ & $q = 0$ \\ 
Static factors with vector autoregressive errors & $n_f = 0$ & $p = 0$ & $q > 0$ \\ 
Static factors & $n_f = 0$ & $p = 0$ & $q = 0$ \\ 
Vector autoregressive errors & $n_f = 0$ & $p = 0$ & $q > 0$ \\ 
Seemingly unrelated regression & $n_f = 0$ & $p = 0$ & $q = 0$ \\ \hline
\end{tabular}
\end{table}
\FloatBarrier
The specified DFM is expressed in state-space form and is estimated by maximum likelihood. The absolute value of the magnitude of the z-statistic associated to the maximum likelihood estimates of the DFMs will be used to evaluate the contribution of each variable as an assessment of the risk in a balance of payments crisis. 

It is essential to clarify that, within the DFM framework, risk is conceptualized as the trend in the latent factor estimated through a Dynamic Factor Model (DFM). This approach allows for the assessment of risk as a systematic and persistent directional movement over time, where an upward trend in the latent factor signals increasing macroeconomic risk. The latent factor represents an aggregate measure of various underlying indicators, enabling the capture of complex, interrelated dynamics in the macroeconomic environment.

This definition contrasts with traditional interpretations of risk, such as volatility or the percentile-based approach. Volatility reflects the dispersion or variability of a variable, providing insight into the degree of uncertainty or fluctuation over a given period. Although informative for understanding short-term risk fluctuations, volatility does not convey the directional or cumulative change inherent in macroeconomic trends captured by the DFM latent factor.

Defining risk in terms of percentiles places a specific value within a historical distribution, emphasizing the extremity or rarity of a particular observation. Although this percentile-based approach effectively identifies outliers or tail risks, it remains a static measure that does not track how risk evolves over time. In contrast, the dynamic latent factor integrates both the magnitude and the directional shifts in risk, offering a forward-looking perspective essential for long-term strategic assessments and policy formulation in macroeconomic analysis. See Appendix A for mathematical details.

Dynamic Factor Models (DFMs) provide a powerful tool for analyzing the Laos PDR economic crisis because they capture the complex interactions between multiple economic indicators over time. Specifically, DFMs enable researchers to:

\begin{itemize}
    \item Identify Key Economic Drivers: DFMs extract a common factor from a large set of economic variables (e.g. inflation, exchange rates, debt, and imports) that drive most economic fluctuations. By applying a DFM, analysts can determine which underlying factors primarily cause the crisis.
    \item Forecast Economic Indicators: DFMs predict the future trajectory of economic indicators such as inflation, GDP growth, and currency exchange rates. Policymakers in Laos can use these forecasts to anticipate economic developments and implement preemptive measures, such as monetary interventions or debt restructuring.
    \item Monitor Spillover Effects: DFMs track how global economic shocks (e.g., fuel price increases or trade disruptions) spill over into Laos’ economy by modeling their impact on domestic variables like inflation, the exchange rate, and the balance of payments.
    \item Evaluate policies: DFMs assess the effectiveness of policy interventions (such as fiscal stimulus or currency management) by simulating how different policies influence the common factor driving the crisis.
\end{itemize}

DFMs offer a comprehensive approach to analyzing Laos’ interconnected economic challenges by identifying the most influential variables and forecasting future trends to support effective policymaking. While both DFMs and Vector Autoregressive (VAR) models are widely used in macroeconomic analysis, DFMs provide several advantages, particularly when dealing with a large number of variables and complex underlying structures.

DFMs excel in reducing the dimensionality of large macroeconomic datasets. It assumes that a latent factor drives the comovement of multiple variables, allowing analysts to model the common structure in the data without estimating an excessive number of parameters. In contrast, VAR models include all observed variables directly, causing the number of parameters to grow quadratically as the number of variables increases. For example, a VAR model with $N$ variables and lag length p requires estimating $N^2p$ parameters. As $N$ grows, VAR models become computationally demanding and increasingly prone to overfitting, especially when working with limited data.

DFMs capture a latent economic factor that drives the dynamics of multiple observed variables. This latent factor often represents key economic conditions such as business cycles, inflationary trends, or financial shocks—forces that are not directly observable but influence the entire economy. By focusing on this common factor, DFMs provide a clearer understanding of the underlying drivers of macroeconomic fluctuations and crises. In contrast, VAR models do not explicitly account for latent factors. Instead, they treat all variables as endogenous and explain each one using the lagged values of all others. This approach can obscure the common forces affecting multiple variables, particularly in high-dimensional settings where unobserved factors drive the true underlying structure.

DFMs offer greater scalability than VAR models when analyzing a large number of time series. By reducing the dataset’s dimensionality and modeling only a few common latent factors, DFMs require estimating only a small number of parameters for each factor and its loadings. This computational efficiency makes DFMs practical even for datasets with many observed variables. In contrast, VAR models become computationally demanding as the number of variables increases because their parameter count grows quadratically. This limitation makes VAR models impractical for large datasets (e.g., more than 10 or 20 variables), leading to potential overfitting and weaker forecasting performance.

DFMs also provide a clearer economic interpretation by extracting common factors that drive the co-movement of multiple variables. For example, a single latent factor could represent the overall business cycle, inflationary pressures, aggregate demand, or macroeconomic risks. This interpretability makes DFMs particularly useful for macroeconomic analysis. In contrast, VAR models treat all variables as endogenous and attempt to explain each using lagged values of itself and others. While this approach allows for complex interactions, it makes it difficult to determine which variables drive the system or identify common economic forces, especially when dealing with many variables.

DFMs tend to outperform VAR models in out-of-sample forecasting, particularly in high-dimensional settings. By filtering out idiosyncratic noise and focusing on key economic drivers, DFMs avoid overfitting and capture long-term trends more effectively. While VAR models can perform well in smaller systems, their forecasting accuracy tends to decline in larger systems due to the high number of parameters. Overfitting becomes a serious issue, especially when handling many variables with limited time-series data.

Finally, DFMs excel at capturing business cycles, inflation trends, macroeconomic risks, and other large-scale economic movements that simultaneously affect multiple variables. The latent factors in DFMs represent these economic trends, making them ideal for analyzing macroeconomic dynamics and crises. While VAR models can capture short-term interrelationships between variables, they lack an explicit structure for identifying broader economic risks or cycles unless augmented with additional constraints or modifications.

\section{Results}

\subsection{Why did the Kip collapse in 2022?}

Table \ref{tab:XR} presents the results of a Dynamic Factor Model (DFM) applied to key annual macroeconomic variables to analyze the exchange rate crisis in Laos PDR. The model includes foreign direct investment (FDI), inflation (INF), external debt (long-term debt), public debt service (DS), total reserves as a percentage of total debt (RPD), GDP growth (GDPG), current account balance (CAB), and exchange rates with major currencies (the Thai baht: XRTHB, and the Chinese renminbi: XRCNY). To ensure comparability, we standardized these variables to have a mean of zero and a variance of one.

\begin{figure}
    \centering
    \caption{Laos: Foreign direct investment (net, BoP, USD)}
    \includegraphics[width=\linewidth]{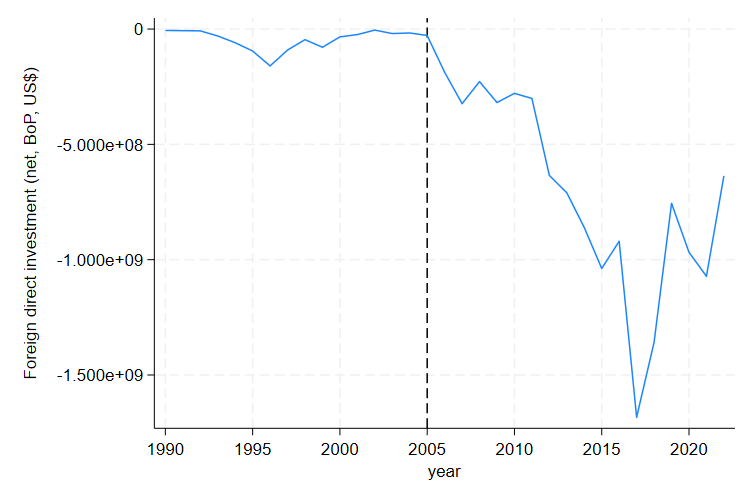}
    \label{fig:FDI}
\end{figure}

\begin{figure}
    \centering
    \caption{Laos: Total reserves (\% of total debt)}
    \includegraphics[width=\linewidth]{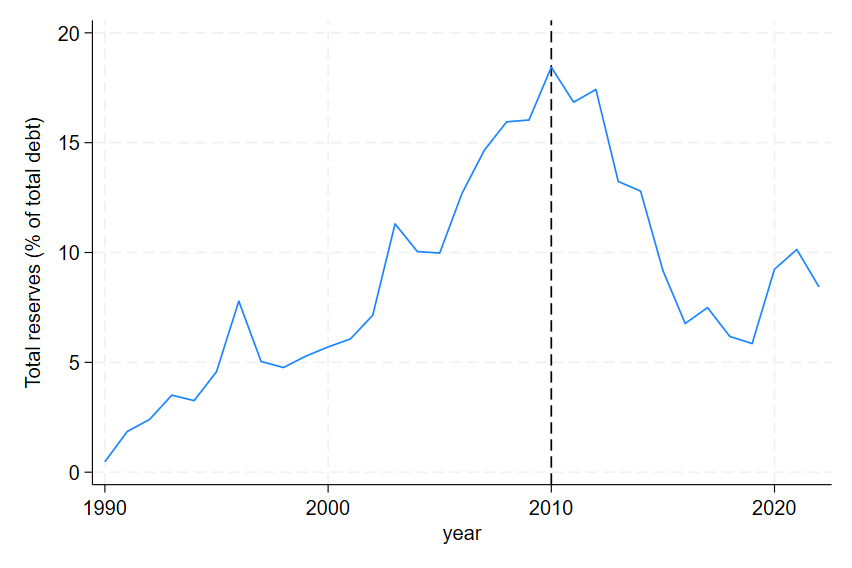}
    \label{fig:RPD}
\end{figure}

\begin{figure}
    \centering
    \caption{Laos: Current account balance (BoP, USD)}
    \includegraphics[width=\linewidth]{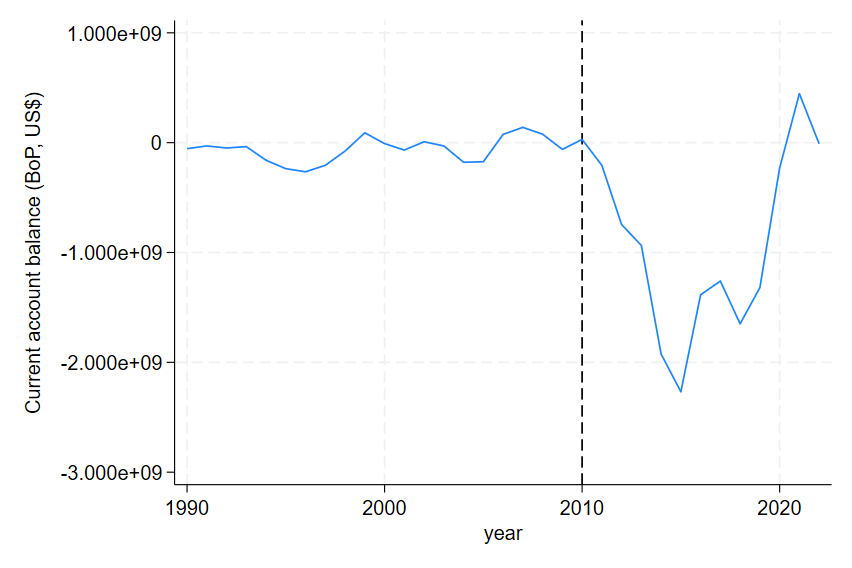}
    \label{fig:CAB}
\end{figure}

\begin{figure}
    \centering
    \caption{Laos: Public and publicly debt service (\% GNI)}
    \includegraphics[width=\linewidth]{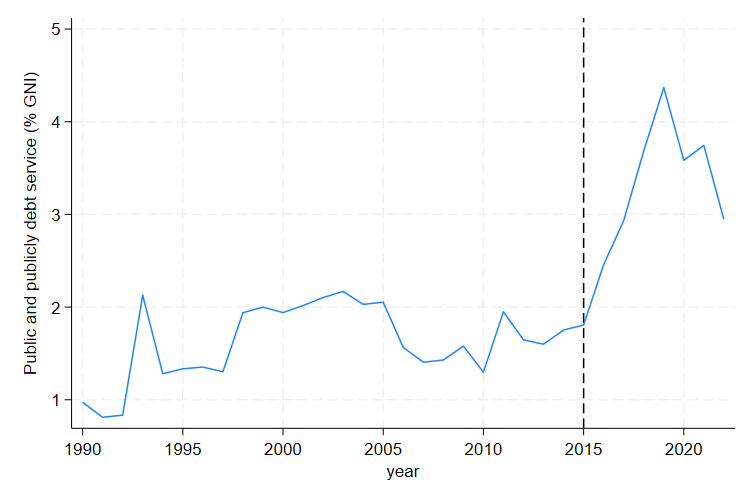}
    \label{fig:DS}
\end{figure}

The model results show that the decline in foreign direct investment, which began in 2005 (Figure \ref{fig:FDI}), played a major role in the exchange rate devaluation of the Kip. The deterioration of the current account balance and the reduction in total reserves (as a percentage of debt) since 2010 (Figures \ref{fig:CAB} and \ref{fig:RPD}) also contributed significantly. Additionally, the increase in debt service payments since 2015 (Figure \ref{fig:DS}) further exacerbated the crisis. These findings indicate that Laos’ economic crisis had already begun, as reflected in the balance of payments accounts, even before debt servicing obligations increased and the kip experienced substantial depreciation against the U.S. dollar.

\begin{table}[ht]
    \centering
    \caption{DFM for the exchange rate in Laos (annual data)}
    \includegraphics[width=12cm]{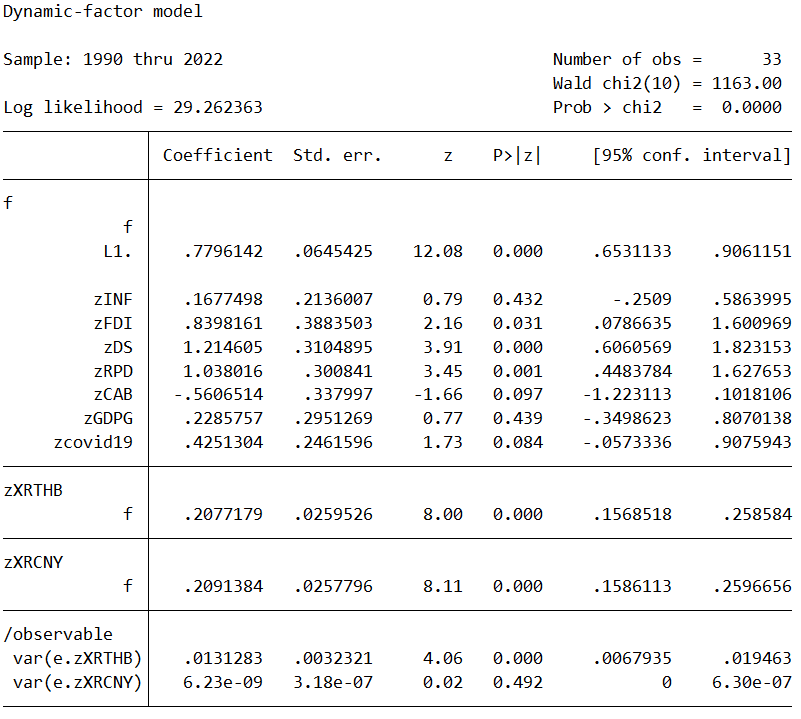}   
    \label{tab:XR}
\end{table}

Table \ref{tab:XRQ} presents an additional model for the exchange rate of the Kip against the USD, estimated using quarterly data. The model results suggest that, in the short term, the worsening balance of payments (BoP) and the increased issuance of narrow money (M1) since 2020 intensified inflationary pressures, created uncertainty, and further depreciated the kip, accelerating its exchange rate decline. However, interpreting these results requires caution due to the limited time window and the level of noise in the quarterly data.

\begin{table}[ht]
    \centering
    \caption{DFM for the exchange rate in Laos (quarterly data)}
    \includegraphics[width=12cm]{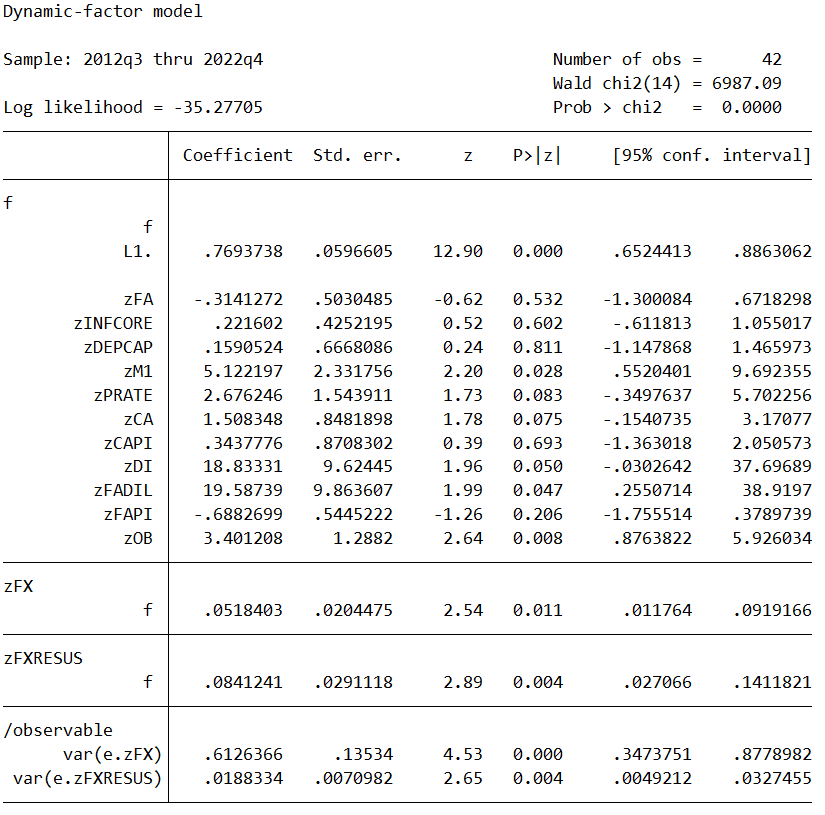}   
    \label{tab:XRQ}
\end{table}

\FloatBarrier

\begin{figure}
    \centering
    \caption{Latent risk factor for the exchange rate in Laos: long-term}
    \includegraphics[width=\linewidth]{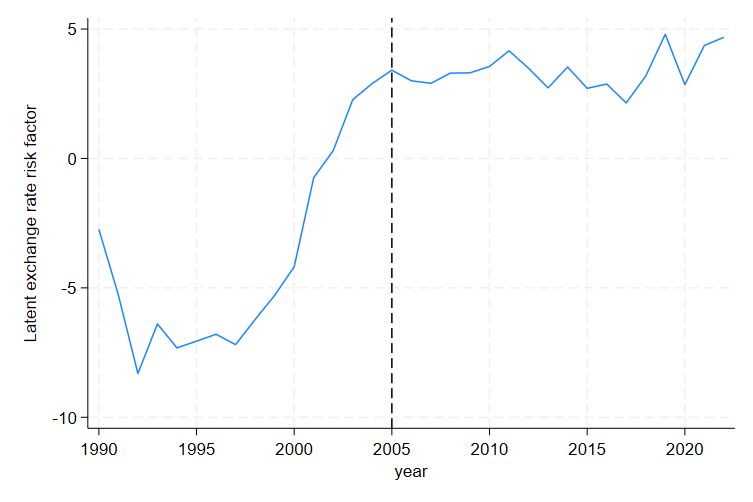}
    \label{fig:XRL}
\end{figure}

\begin{figure}
    \centering
    \caption{Latent risk factor for the exchange rate in Laos: short-term}
    \includegraphics[width=\linewidth]{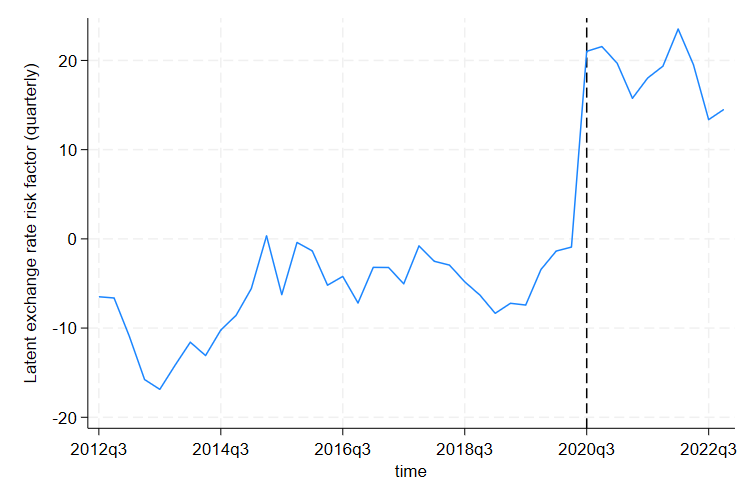}
    \label{fig:XRLQ}
\end{figure}

\begin{figure}
    \centering
    \caption{Laos: Narrow Money (M1, LAK)}
    \includegraphics[width=\linewidth]{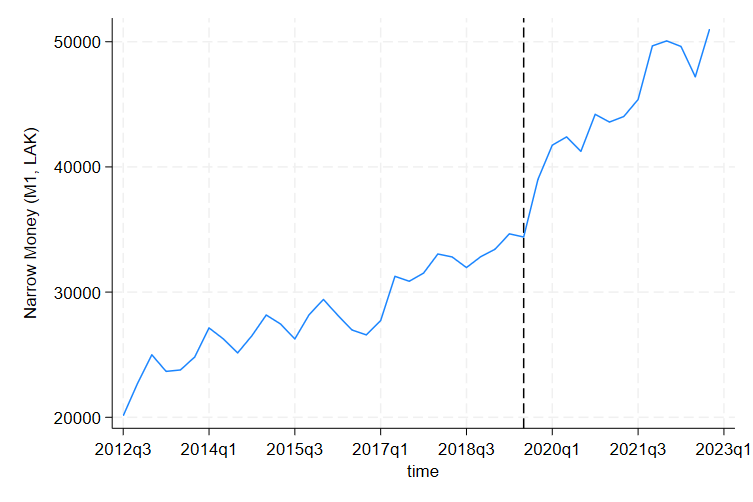}
    \label{fig:M1}
\end{figure}

\begin{figure}
    \centering
    \caption{Laos: Overall balance (BoP)}
    \includegraphics[width=\linewidth]{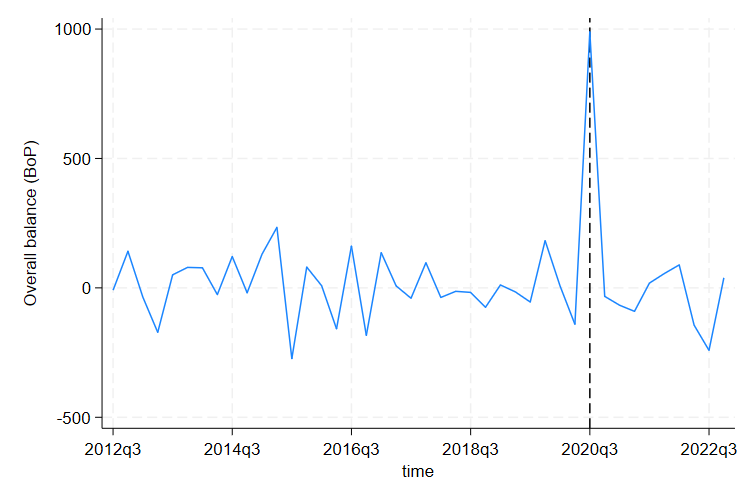}
    \label{fig:OB}
\end{figure}

\FloatBarrier

Statistically, the results of the DFM can be interpreted as follows:

\begin{itemize}
    \item \textbf{Factor loading and lagged factor (f, L1)}: The estimated coefficient (0.7796) is significantly positive and large (z = 12.08, p-value = 0.000), indicating that the latent risk factor captures exchange rate pressures and related macroeconomic instability over time. The persistence of the latent factor suggests that external shocks to the exchange rate, such as depreciation pressures, have long-lasting effects and highlight structural issues reflected in balance of payments movements. This finding implies that Laos’ economic structure remains fundamentally weak, making it more vulnerable to prolonged shocks compared to a country with a diversified and strong economy and a healthier balance of payments, particularly in the capital and financial accounts.
    \item \textbf{Foreign Direct Investment (zFDI):} The coefficient for FDI (0.8398) is positive and statistically significant at the 5\% but not at the 1\% level (z = 2.16, p-value = 0.031). \textbf{This indicates that an increase in FDI is positively associated with the latent factor.} In the context of the exchange rate crisis, this finding suggests that while foreign direct investment can provide some stability, it does not fully offset the broader structural issues affecting the exchange rate, such as the rising cost of servicing external debt in Laos.
    \item \textbf{Public Debt Service (zDS):} The coefficient for public debt service (1.2146) is positive and highly significant (z = 3.91, p-value = 0.000). This result indicates that rising debt service obligations, as a percentage of GNI, heavily drive the latent factor behind the exchange rate crisis. The increasing debt burden intensifies pressure on foreign exchange reserves, devalues the Kip, and makes it more difficult for Laos to manage its external debt.
    \item \textbf{Total Reserves to Debt Ratio (zRPD):} The coefficient (1.038) is positive and significant at the 1\% level (z = 3.45, p-value = 0.001). This result shows that a higher reserve-to-debt ratio increases pressure on the exchange rate, suggesting that as Laos depletes its reserves to service debt or defend the currency, economic instability worsens. It underscores the vulnerability of Laos’ reserves amid high debt levels.
    \item \textbf{Current Account Balance (zCAB):} The coefficient for the current account balance is negative (-0.5607) and statistically significant only at the 10\% level (p-value = 0.097). The negative sign suggests that a worsening current account balance, which reflects higher imports relative to exports, may be helping elevate exchange rate pressures. 
    \item \textbf{GDP Growth (zGDPG):} The coefficient for GDP growth (0.2288) is positive, but not statistically significant at conventional level (p-value = 0.439). This result indicates that while economic growth usually contributes to stability, it did not play a crucial role in reducing latent exchange rate pressures in Laos during this period. Despite economic growth, structural vulnerabilities have persisted.
    \item \textbf{COVID-19 Impact (zcovid19)}: The COVID-19 coefficient (0.4251) is positive but only significant at the 10\% level (p-value = 0.084). This result indicates that the pandemic affected the economy by intensifying exchange rate pressures, disrupting economic activity, reducing foreign exchange earnings, and increasing reliance on external debt to balance the budget. However, it was not the primary driver of the exchange rate crisis in Laos.  
\end{itemize}

\textbf{The latent exchange rate risk factor over time (Figure \ref{fig:XRL}) shows a growing trend in exchange rate risk over the years, peaking around the 2020 period, likely due to the impacts of the COVID-19 pandemic and debt issues. }The persistent upward trajectory indicates that external risks have increasingly destabilized Laos’ currency. The sharp increase around 2020 aligns with global economic disruptions caused by the pandemic, but the long-term trend reveals that these problems have been increasing for decades. External debt and limited reserves have mainly driven this trend, as these two variables exhibit the largest z-statistic values in the maximum likelihood estimates of the DFM, making them the two main contributors to Laos’ exchange rate crisis. Although the statistical results identify debt servicing costs as the main driver of the crisis (z-statistic = $|3.91|$), this effect does not act in isolation. Depletion of reserves (z-statistic = $|3.45|$) also played an important role, along with other variables with a lesser impact. The z statistics result from dividing the estimated coefficient by its standard error: for debt servicing costs, $\frac{1.214605}{0.3104895} = 3.9119036$, and for reserves, $\frac{1.038016}{0.300841} = 3.4503808$.

The DFM results highlight that Laos PDR's exchange rate crisis was not driven by a single variable but rather by multiple factors. A combination of high external debt servicing costs, depleting reserves, and dependencies on foreign currencies like the Thai Baht and Chinese Yuan drove Laos' exchange crisis. While FDI provided some stabilization, structural vulnerabilities, such as the current account deficit and public debt burden, continued to pressure the exchange rate. The latent factor model effectively captures these dynamics, illustrating the persistence of exchange rate risk and the critical role macroeconomic variables play in worsening the crisis.

\subsection{Can We Predict a Balance of Payments Crisis with the available data?}

The Dynamic Factor Model (DFM) in Tables \ref{tab:BoP_DFM_g} and \ref{tab:BoP_DFM_r} analyzes key components of Laos’ balance of payments (BoP), including the current account (CA), capital account (KA), financial account (FA), primary income of the current account (CAPI), and portfolio investment (FAPI). Table \ref{tab:BoP_DFM_g} presents a general (saturated) model that includes all variables, while Table \ref{tab:BoP_DFM_r} shows a reduced model that retains only statistically significant variables at the 5

\begin{figure}[hb]
    \centering
    \caption{Latent risk factor for the balance of payments in Laos}
    \includegraphics[width=0.97\linewidth]{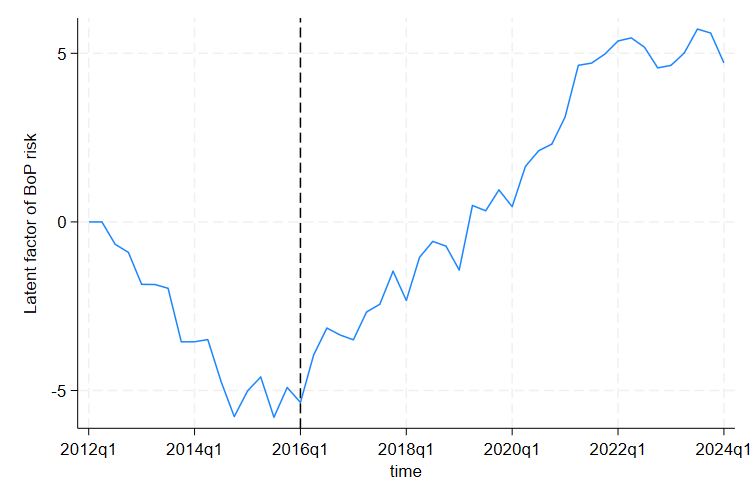}
    \label{fig:BoPL}
\end{figure}

\begin{figure}[hb]
    \centering
    \caption{Laos: Current account primary income (USD)}
    \includegraphics[width=0.97\linewidth]{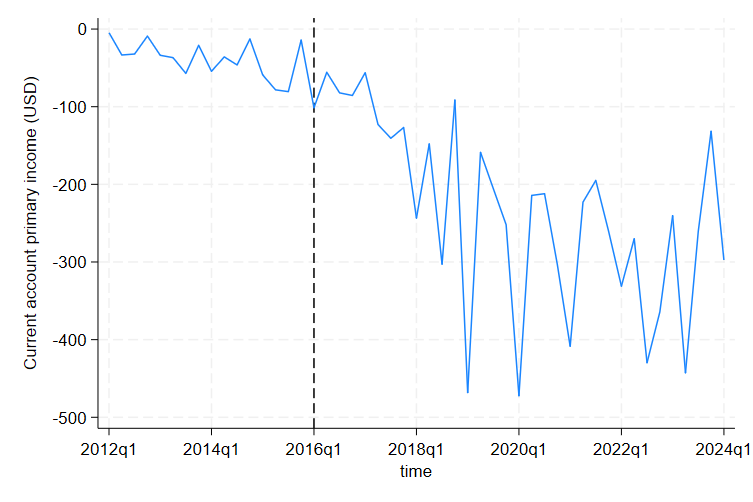}
    \label{fig:CAPI_BoP}
\end{figure}

\begin{figure}[hb]
    \centering
    \caption{Laos: Capital account (USD)}
    \includegraphics[width=0.97\linewidth]{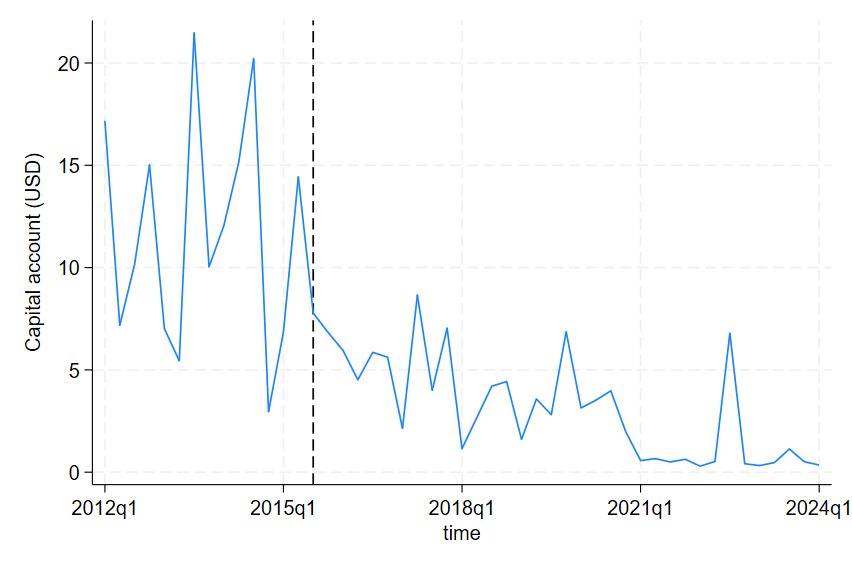}
    \label{fig:KA_BoP}
\end{figure}

This key BoP factor began deteriorating in 2016 (Figure \ref{fig:BoPL}), driven primarily by the worsening of the current account’s primary income (Figure \ref{fig:CAPI_BoP}) and the capital account (Figure \ref{fig:KA_BoP}) from that point onward. The decline would have served as an early warning for a potential external sector crisis, providing policymakers with valuable insights to implement timely measures and prevent further instability.

\begin{itemize}
\item Lagged Factor (f, L1): The coefficient for the lagged factor (0.9722) is large and highly significant (z = 33.09, p-value = 0.000). This result indicates that the latent factor driving BoP risk remains highly persistent over time, meaning that once imbalances emerge in the balance of payments, they tend to persist or even worsen in subsequent periods.

\item Current Account (zCA): The coefficient for the current account (0.1602) is positive and significant (z = 3.63, p-value = 0.000). A stronger current account surplus helps reduce the latent risk associated with the balance of payments, while deficits worsen the BoP situation and prolong economic shocks. Since the current account reflects trade balances and income flows, its deterioration serves as an early warning indicator of a potential BoP crisis.

\item Capital Account (zKA): The coefficient for the capital account (-0.1643) is negative and statistically significant at the 1

\item Financial Account (zFA): The financial account coefficient (0.1236) is positive and statistically significant at the 1

\item Current Account Primary Income (zCAPI): The coefficient for current account primary income (-0.1812) is negative and statistically significant at the 1

\item Portfolio Investment (zFAPI): The coefficient for portfolio investment (0.0863) is positive and statistically significant at the 5
\end{itemize}

The graph of the latent factor of BoP risk (Figure \ref{fig:BoPL}) tells an interesting story with its variations. BoP risks decline until 2016, likely due to improvements in the current and financial accounts during that period. However, after 2016, the risk factor steadily rises, peaking around 2022–2023. This period corresponds with increasing external vulnerabilities exacerbated by external shocks such as the COVID-19 pandemic, global financial volatility, and mounting external debt pressures.

\begin{table}[ht]
    \centering
    \caption{General DFM for the balance of payments in Laos}
\includegraphics[width=12cm]{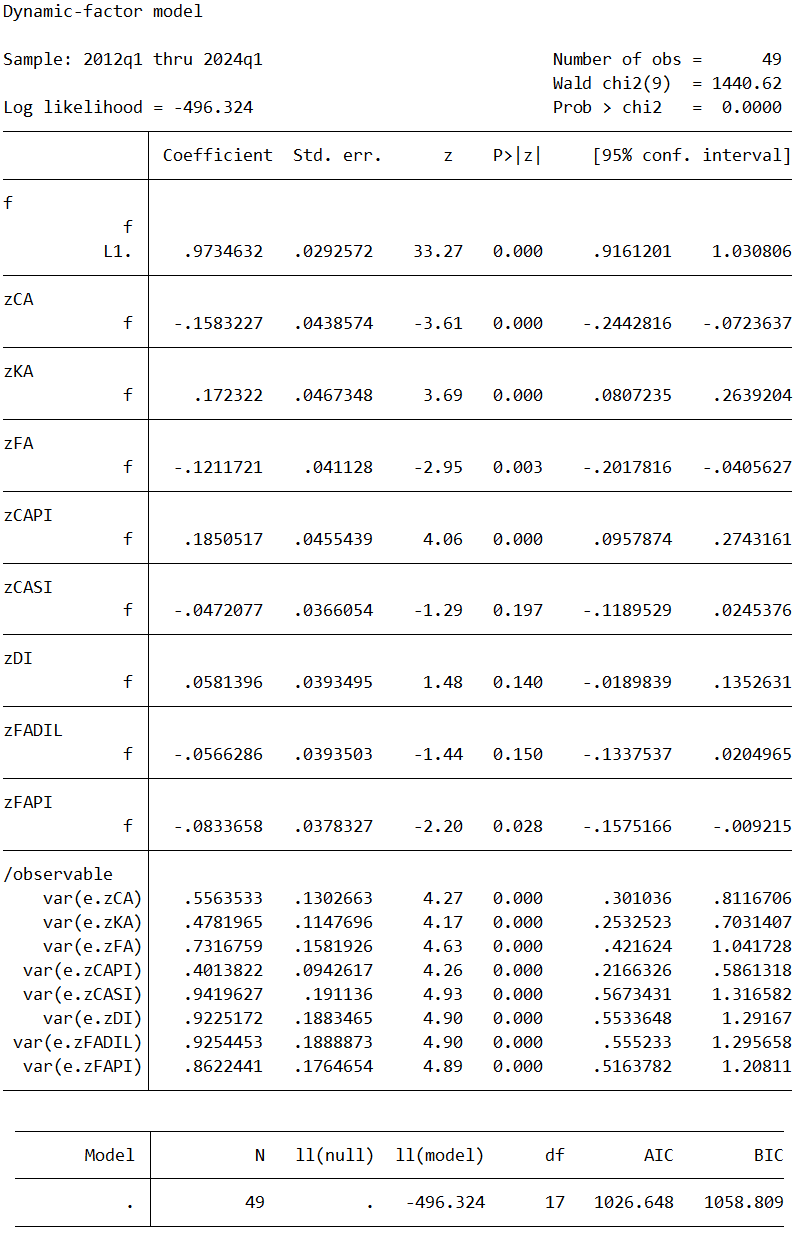}   
     \label{tab:BoP_DFM_g}
\end{table}

\begin{table}[ht]
    \centering
    \caption{Reduced DFM for the balance of payments in Laos}
\includegraphics[width=12cm]{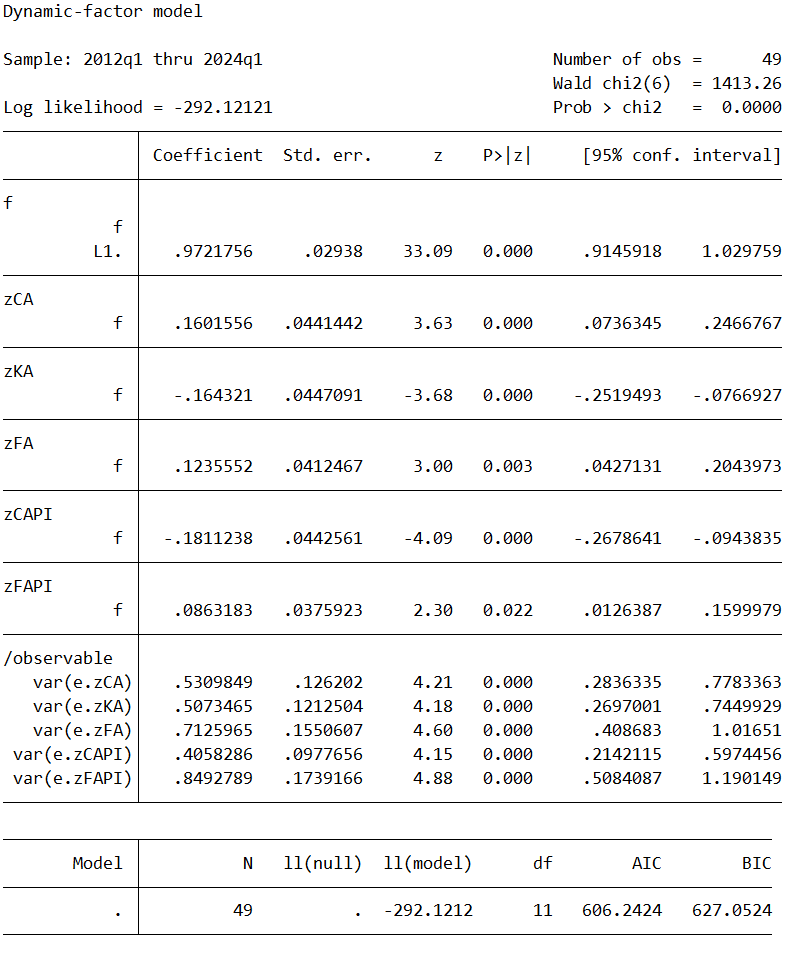}   
     \label{tab:BoP_DFM_r}
\end{table}
\FloatBarrier

A balance of payments (BoP) crisis typically emerges due to a major current account deficit, capital flight, or a financial account crisis. Monitoring and projecting the evolution of the latent risk factor $f_t$ can help predict such crises. The following approaches provide predictive insights:

\begin{itemize}
\item Monitoring sharp increases in $f_t$: A sudden rise in $f_t$ signals growing instability, making it a valuable early warning indicator. The DFM results highlight the persistence of BoP imbalances. The strong autocorrelation in the latent factor (f, L1) shows that once vulnerabilities emerge, they continue to persist. Detecting rising BoP risk early through $f_t$ allows policymakers to intervene before a full crisis unfolds.

\item Tracking capital outflows and other BoP risk drivers: Policymakers can predict a major capital account crisis by monitoring capital flow data, particularly large and sudden foreign investment outflows. The DFM identifies key contributors to BoP risk, such as the capital account, financial account, and primary income. Monitoring these components enables targeted measures, such as encouraging capital inflows or strengthening reserves, to mitigate risks before they escalate.

\item Identifying financial account vulnerabilities: If a significant portion of external debt is short-term and denominated in foreign currency, a sudden exchange rate depreciation could trigger a financial account crisis. Sharp increases in $f_t$ would signal stress in the economy’s ability to manage external debt obligations, helping policymakers take preventive actions.

\item Conducting scenario analysis: The DFM allows for scenario simulations to assess how specific changes in accounts—such as a worsening current account or an improvement in portfolio investment—affect the latent BoP risk factor. Stress-testing the economy against external shocks, including capital flight, declining FDI, or deteriorating terms of trade, can inform more effective policy responses.

\item Forecasting future crises: The DFM can forecast the latent BoP risk factor over future quarters. A steady increase in $f_t$ signals an impending BoP crisis, allowing policymakers to take preemptive measures such as tightening monetary policy, negotiating debt restructuring, or seeking external financial assistance.
\end{itemize}

\section{Conclusion}

This document applied Dynamic Factor Models (DFMs) to examine two critical macroeconomic issues: the causes of the exchange rate collapse in 2022 and the prediction of balance of payments crises. The DFM results for Laos PDR’s exchange rate indicate that depreciation resulted from multiple factors, including declining foreign direct investment since 2005, a worsening current account balance and depleting reserves since 2010, and rising external debt servicing costs since 2015. Additionally, the DFM provided key insights into the drivers of BoP risk in Laos, identifying capital account deterioration and declining primary income as major contributors to a BoP crisis.

\appendix
\section*{Appendix A: Risk Concepts}

\subsection*{A1. Risk as a Trend in a Latent Factor}
Risk as a trend can be modeled using a dynamic factor model to capture the evolution of a latent factor over time. This is particularly useful for assessing systemic risk across multiple variables. A dynamic factor model can be represented as:

\begin{align}
X_{t} &= \Lambda F_{t} + \epsilon_{t}, \\
F_{t} &= A F_{t-1} + \eta_{t},
\end{align}

where:
\begin{itemize}
    \item $X_{t}$ is the vector of observed variables at time $t$,
    \item $\Lambda$ is the loading matrix that maps the latent factors to the observed variables,
    \item $F_{t}$ is the vector of latent dynamic factors at time $t$,
    \item $\epsilon_{t}$ is the vector of idiosyncratic errors, assumed to be white noise,
    \item $A$ is the transition matrix describing the dynamics of the latent factors,
    \item $\eta_{t}$ is the vector of process noise, assumed to be white noise.
\end{itemize}

The estimation of the latent dynamic factor $F_{t}$ can be performed using the Kalman filter, which recursively estimates the state of a linear dynamic system:

    \begin{equation}
    \hat{F}_{t|t-1} = A \hat{F}_{t-1|t-1},
    \end{equation}
    \begin{equation}
    P_{t|t-1} = A P_{t-1|t-1} A^\top + Q,
    \end{equation}
    where $P_{t|t-1}$ is the predicted covariance matrix and $Q$ is the covariance matrix of $\eta_{t}$.
    
    \begin{equation}
    K_{t} = P_{t|t-1} \Lambda^\top (\Lambda P_{t|t-1} \Lambda^\top + R)^{-1},
    \end{equation}
    \begin{equation}
    \hat{F}_{t|t} = \hat{F}_{t|t-1} + K_{t} (X_{t} - \Lambda \hat{F}_{t|t-1}),
    \end{equation}
    \begin{equation}
    P_{t|t} = (I - K_{t} \Lambda) P_{t|t-1},
    \end{equation}
    where $K_{t}$ is the Kalman gain matrix, $R$ is the covariance matrix of $\epsilon_{t}$, and $I$ is the identity matrix.

The Kalman filter iteratively estimates $F_{t}$, providing a dynamic assessment of the latent factor that reflects the trend in the risk over time.

\subsection*{A2. Risk as Volatility}
Risk volatility measures the extent of variation or dispersion of a risk factor over time. It is defined as the standard deviation of a time series $X_t$:

\begin{equation}
\sigma = \sqrt{\frac{1}{N} \sum_{t=1}^{N} (X_t - \mu)^2},
\end{equation}

where:
\begin{itemize}
    \item $\mu$ is the mean of the time series $X_t$,
    \item $N$ is the number of observations.
\end{itemize}

Volatility reflects the degree to which $X_t$ deviates from its average, indicating how much risk fluctuates over time.

\subsection*{A3. Risk as the Percentile of a Distribution}
A percentile provides a measure of where a particular value lies within a distribution relative to the entire dataset. The $p$-th percentile $P(X, p)$ is defined as:

\begin{equation}
P(X, p) = X_{\lceil p(N+1) \rceil},
\end{equation}

where:
\begin{itemize}
    \item $p \in (0, 1)$ is the percentile (e.g., $p = 0.75$ for the 75th percentile),
    \item $\lceil \cdot \rceil$ denotes the ceiling function to find the appropriate index in the ordered set.
\end{itemize}

The percentile indicates the value below which $p \times 100\%$ of the observations fall. For instance, if a risk factor's value lies at the 90th percentile, it is higher than 90\% of historical values, indicating it is in the upper tail of the distribution and may be considered high.

\bibliography{refs}

\end{document}